\documentclass[aps,prl,preprint,showpacs]{revtex4-1}
\usepackage{graphicx}
\usepackage{latexsym}
\usepackage[T1]{fontenc}
\usepackage{endnotes}
\usepackage{lmodern}
\input{epsf}

\begin{document}

\preprint{APS/123-QED}

\date{\today}

\title{Superconductivity-induced reentrance of the orthorhombic distortion in Ba$_{1-x}$K$_{x}$Fe$_2$As$_2$}

\author{A. E. Böhmer}
\email{present address: Ames Laboratory/ Iowa State University, boehmer@ameslab.gov}
\affiliation{Institut für Festkörperphysik, Karlsruhe Institute of Technology, 76021 Karlsruhe, Germany}

\author{F. Hardy} 
\affiliation{Institut für Festkörperphysik, Karlsruhe Institute of Technology, 76021 Karlsruhe, Germany}

\author{L. Wang} 
\affiliation{Institut für Festkörperphysik, Karlsruhe Institute of Technology, 76021 Karlsruhe, Germany}

\author{T. Wolf}
\affiliation{Institut für Festkörperphysik, Karlsruhe Institute of Technology, 76021 Karlsruhe, Germany}

\author{P. Schweiss}
\affiliation{Institut für Festkörperphysik, Karlsruhe Institute of Technology, 76021 Karlsruhe, Germany}

\author{C. Meingast}
\email{christoph.meingast@kit.edu}
\affiliation{Institut für Festkörperphysik, Karlsruhe Institute of Technology, 76021 Karlsruhe, Germany}

\begin{abstract}
Detailed knowledge of the phase diagram and the nature of the competing magnetic and superconducting phases is imperative for an understanding of the physics of iron-based superconductivity.  Here, we show using thermodynamic probes that the phase diagram of the first discovered, and highest $T_c$, '122'-type material, Ba$_{1-x}$K$_{x}$Fe$_2$As$_2$ is in fact much richer than previously reported. Inside the usual stripe-type magnetic order with $C_2$-symmetry, we find a small pocket of a tetragonal, $C_4$-symmetric phase, which surprisingly reverts back to the $C_2$-phase at or slightly below the superconducting transition. This re-entrance to a low-temperature orthorhombic state induced by superconductivity is discussed in terms of competition of the two magnetic phases with superconductivity and is illustrated by the measured changes in the electronic entropy of the system. Using our thermodynamic data, we make predictions about how the phase diagram of these competing orders will change under pressure.  
\end{abstract}


\maketitle

Unconventional superconductivity often arises around the point where some kind of magnetic order (either antiferromagnetic (AFM) or ferromagnetic (FM)) is suppressed by a tuning parameter, such as pressure or chemical substitution \cite{Flouquet2005,Pfleiderer2009,Norman2011}. This led to the idea that superconducting pairing in these materials may result from the low-energy magnetic fluctuations surrounding the quantum critical point where magnetism is suppressed to zero temperature \cite{Marthur1998}. The magnetic ground state of typical iron-based superconducting parent compounds is a stripe-type AFM spin-density wave (SDW) phase, which breaks the $C_4$ symmetry of the high-temperature tetragonal paramagnetic phase and is closely related to a tetragonal-to-orthorhombic structural distortion of the lattice \cite{Paglione2010}. An important debate, triggered by the observation that this orthorhombic structural distortion sometimes occurs slightly above the magnetic transition \cite{Canfield2010}, concerns the respective roles of spin and orbital degrees of freedom \cite{Fernandes2014,Kontani2011}. In the orbital picture, the structural transition is due to orbital ordering and can trigger a, secondary, magnetic transition \cite{Kontani2011}. In the spin scenario, on the other hand, magnetism is essential and the structural transition is really just the first step of the magnetic transition, induced by an Ising-nematic degree of freedom associated with the emerging stripe-type magnetic order \cite{Fernandes2012}. Importantly, both orbital and spin fluctuations are candidates for the superconducting pairing \cite{Fernandes2014,Kontani2011}.

A surprising recent result is the observation of a $C_4$-symmetric tetragonal magnetically ordered phase in (Na,Ba)Fe$_2$As$_2$ using neutron scattering \cite{Avci2014}, which was taken as evidence for the spin scenario because the existence of such a phase can hardly be reconciled with orbital order being a prerequisite for magnetism \cite{Avci2014}. More recently, however, single crystal neutron scattering has provided evidence for a spin reorientation in the $C_4$ phase \cite{Wasser2014}, demonstrating that orbital physics and spin-orbit coupling cannot be neglected. 
Up to now, a $C_4$-magnetic phase has not been observed in the closely related Ba$_{1-x}$K$_{x}$Fe$_2$As$_2$ system, which was the first iron-based superconductor of 122-type stoichiometry to be discovered \cite{Rotter2008} and has the highest $T_c=38$ K within this family. The observation of an unidentified phase transition in resistivity measurements under pressure \cite{Hassinger2012}, however, hints that an additional instability is close by. 

Here, we re-examine the phase diagram of Ba$_{1-x}$K$_{x}$Fe$_2$As$_2$ in unprecedented detail using thermal-expansion and specific-heat measurements of very high-quality single crystals. Within a narrow composition range, we observe a $C_4$-symmetric magnetic phase, which apparently has been missed in previous investigations \cite{Avci2011,Avci2012,Hassinger2012,Tanatar2014}. In strong contrast to Na-doped BaFe$_2$As$_2$, the $C_4$-phase is, however, not stable to zero temperature and reverts back to the $C_2$ phase in the vicinity of the onset of superconductivity. This apparent preference of superconductivity for the stripe-type $C_2$ over the $C_4$ magnetic state in this system may provide important clues, not only concerning the spin-orbital debate, but also about the superconducting pairing mechanism itself.\\

\begin{figure*}
\includegraphics[width=\textwidth]{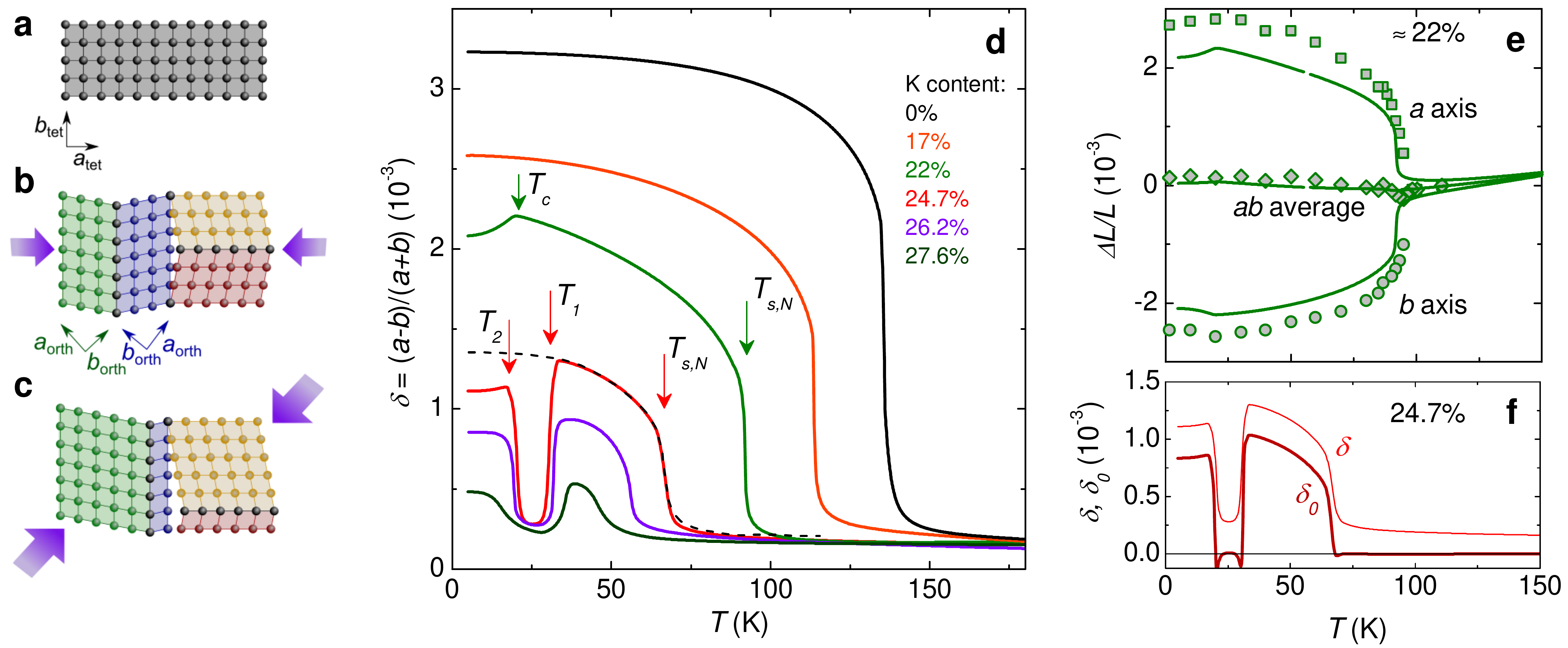}
\caption{\small {Orthorhombic distortion of Ba$_{1-x}$K$_{x}$Fe$_2$As$_2$ as measured using a capacitance dilatometer. (a) Schematic representation of the tetragonal $C_4$ high-temperature phase. (b) Samples in the orthorhombic $C_2$ phase form structural domains ('twins'), which are unaffected by the force from the spring-loaded dilatometer if it is applied along the tetragonal [100] in-plane direction (purple arrows). (c) If the dilatometer force is applied along [110], domains with their (shorter) orthorhombic $b$ axis along the direction of the force are selected ('detwinning'). (d) Temperature dependence of the orthorhombic distortion $\delta = (a - b)/(a + b)$ of underdoped Ba$_{1-x}$K$_{x}$Fe$_2$As$_2$ obtained using difference of 'twinned' and 'detwinned' data from a high-resolution capacitance dilatometer (see Fig. 1 a-c). Abrupt changes of $\delta$ mark structural phase transitions. (e) Good agreement between our results (continuous lines), and results from neutron powder diffraction \cite{Avci2011} (symbols) for the thermal expansion of the $a$ and $b$ axis demonstrates the reliability of our technique. (f) Orthorhombic distortion corrected for the effect of the applied force, $\delta_0$, for the sample with 24.7\% K content, distinguishing tetragonal and orthorhombic phases.}}
\label{fig:1}
\end{figure*}

Fig. 1 presents our first main result, the orthorhombic distortion $\delta = (a-b)/(a+b)$ ($a$ and $b$ are the in-plane lattice constants) versus temperature of underdoped Ba$_{1-x}$K$_{x}$Fe$_2$As$_2$ obtained using a high-resolution capacitance dilatometer \cite{Meingast1990}. Even though  dilatometry is a macroscopic probe, the thermal expansion of the individual $a$ and $b$ axes, as well as the distortion $\delta$, can be derived using the difference between 'twinned' and 'detwinned' data sets, as shown previously for FeSe \cite{Boehmer2013} and detailed in the supplementary material. The reliability of this method, which has higher resolution by several orders of magnitude than either neutron or x-ray diffraction experiments, is demonstrated by the close match to results of neutron powder diffraction \cite{Avci2011} (see Fig. 1 e). In Ba$_{1-x}$K$_{x}$Fe$_2$As$_2$, the structural and magnetic transitions are coincident and first-order over the entire phase diagram \cite{Avci2011}. Our curves in Fig. 1 d, indeed, exhibit the well-known increase of $\delta$ at $T_{s,N}$ and the subsequent suppression of $\delta$ below $T_c$ \cite{Wiesenmayer2011} for 22\% K content. In the small range $\sim24.7\% - 27.6\%$ K content we find a surprising new result, namely a sudden reduction of the orthorhombic distortion at $T_1<T_{s,N}$, followed by a sudden increase at $T_2$ to a value slightly below the maximum value reached above $T_1$. This sudden reduction of $\delta$ strongly suggests that these Ba$_{1-x}$K$_{x}$Fe$_2$As$_2$ samples undergo a similar transition to a $C_4$ magnetic phase as found in (Ba,Na)Fe$_2$As$_2$\cite{Avci2014,Wasser2014}. We take the increase of $\delta$ at $T_2$ as indication for reentrance of the $C_2$ SDW phase. To prove that the intermediate phase is truly tetragonal, we have corrected the data for the 24.7\% K crystal for the finite stress applied during the measurement (Fig. 1 f), which induces non-zero distortion $\delta$ even in $C_4$-symmetric phases, using the accurately determined elastic modulus of the sample (see supplemental material for details of the procedure). We, indeed, find that the corrected value, $\delta_0$, vanishes completely (to within the extremely small value $\sim 0.01 \times 10^{-3}$, related to our error bar) in the intermediate phase, as expected for truly tetragonal $C_4$ phases.

Similar to inverse melting \cite{Greer2000}, the apparently counterintuitive transition at $T_1$ into a state of seemingly higher symmetry upon lowering of the temperature does not violate the laws of thermodynamics. Rather, it is an indication that the structural distortion is not primary the order parameter and that other, likely magnetic, degrees of freedom must be involved. The absolute measure for disorder is the entropy of the isolated system, which never increases with decreasing temperature. In the following we investigate the novel re-entrant transitions at $T_1$ and $T_2$ and their relation to superconductivity by examining the electronic entropy of the system. 
Moreover, the dotted black line for the 24.7\% sample (red curve in Fig. 1 d) indicates the low-temperature behavior of $\delta$ expected in the absence of the new $C_4$ phase and superconductivity. The fact that $\delta$ below $T_2$ is still reduced from the value expected by extrapolation suggests that the superconducting transition lies somewhere within the intermediate $C_4$ phase. However, no clear signature of superconductivity can be observed in the $\delta(T)$ curves and heat-capacity data are crucial to nail down the location of the bulk superconducting transition.

Fig. 2 presents our electronic heat capacity $C_e/T$ and electronic entropy divided by temperature $S_e/T=\frac{1}{T}\int{C_e/TdT}$, in addition to $\delta$, and the temperature-dependent in-plane length change $\Delta L_{ab}/L_{ab}$ for 'twinned' samples ($c$-axis data and thermal-expansion coefficients are given in the supplementary material) for various compositions. In a Fermi liquid, $S_e/T$ is expected to be constant, and how the entropy is lost upon cooling through the transitions provides important information about the strength of the competing orders. For the 23.2\% sample, the usual first-order tetragonal-to-orthorhombic transition occurs at $T_{s,N}=81$ K and results in a sizable reduction of $S_e/T$. The remaining $S_e/T$ is lost through superconductivity with an onset at $T_c=24$ K (see Fig. 2 a,f,k,p). 

The lowest K concentration for which the $C_4$ magnetic phase is observed is 24.7\% (Fig. 2 b,g,l,q). Surprisingly, we do not observe a clear signature of a superconducting transition neither in $C_e/T$ nor in $L_{ab}$ for this crystal, while strong first-order peaks indicate $T_{s,N}$, $T_1$ and $T_2$ unambiguously. Nevertheless, the heat-capacity data show that the Fermi-surface is completely gapped at low temperature, implying a superconducting ground state. $T_c$ can, however be located by applying a large magnetic field. In a magnetic field of 12 T, the sharp peak at $T_2$ disappears and is replaced by a broadened step-like anomaly, which is slightly shifted downward in temperature from the peak at $T_2$, while the transitions at $T_1$ and $T_{s,N}$ are hardly affected. The shift of the transition and the field broadening is the expected behaviour for a superconducting transition, and we therefore identify this transition with $T_c$. This result implies that the strong specific-heat peak in zero-field at $T_2$ is a novel combined structural, magnetic, and superconducting first-order transition, in which re-entrance of the $C_2$-SDW phase occurs. How exactly a magnetic field tunes the system from a first- to a second-order transition at $T_c$ is a direction for future studies.

\begin{figure*}
\includegraphics[width=\textwidth]{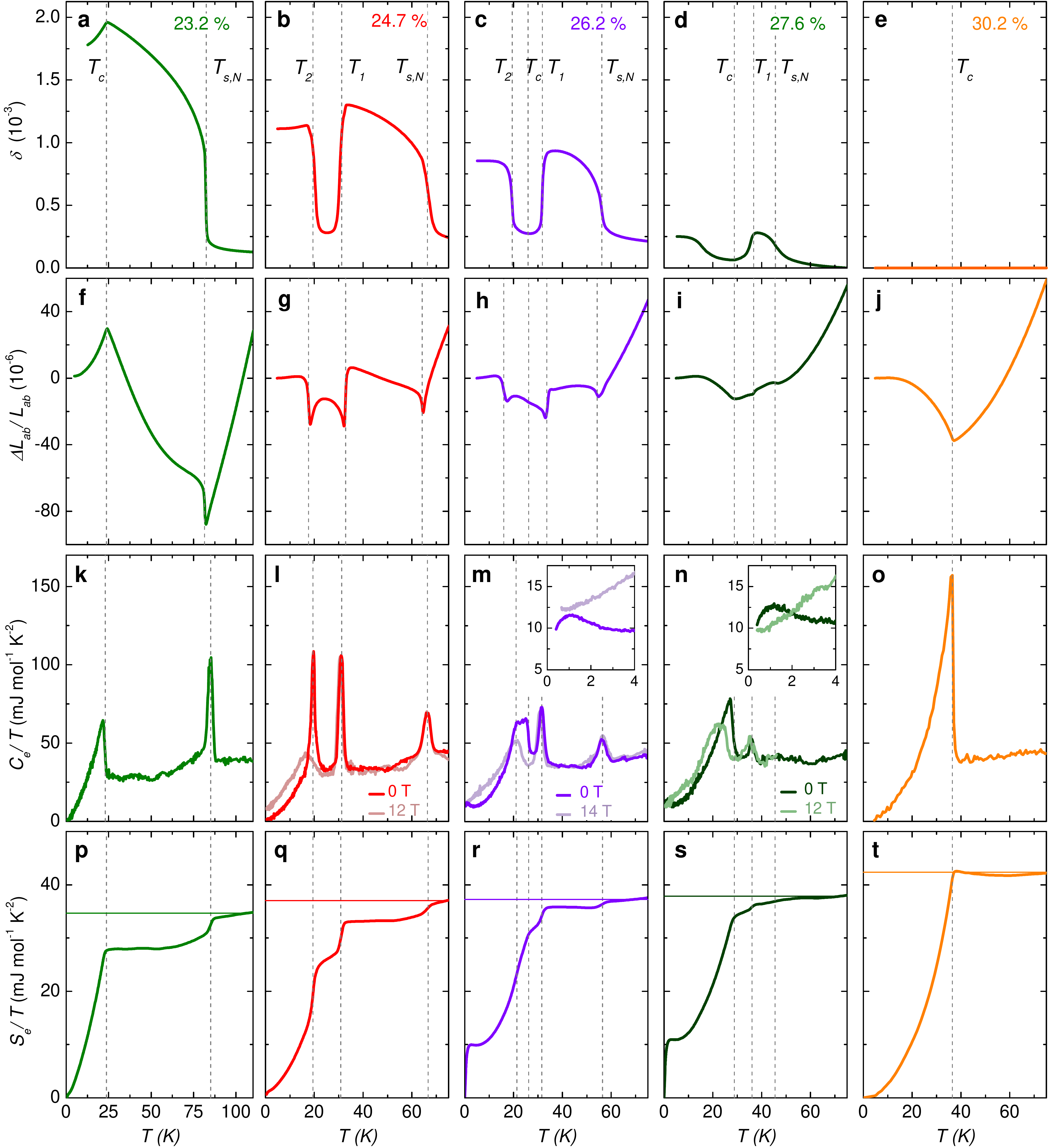}
\caption{\small {Phase transitions in thermal expansion and specific heat for 5 samples of different K content. (a-e) Orthorhombic distortion $\delta$, (f-j) average in-plane length change $\Delta L_{ab}/L_{ab}$ for 'twinned' samples (k-o) electronic specific heat $C_e/T$, (p-t) electronic entropy divided by temperature $S_e/T$. Horizontal lines indicate the value of $S_e/T$ of the high-temperature paramagnetic phase which is lost upon cooling through the various transitions. (l,m,n) also show data measured in a large magnetic field and the insets show a magnification of the low-temperature region.}}
\label{fig:2}
\end{figure*}


\begin{figure*}
\includegraphics[width=\textwidth]{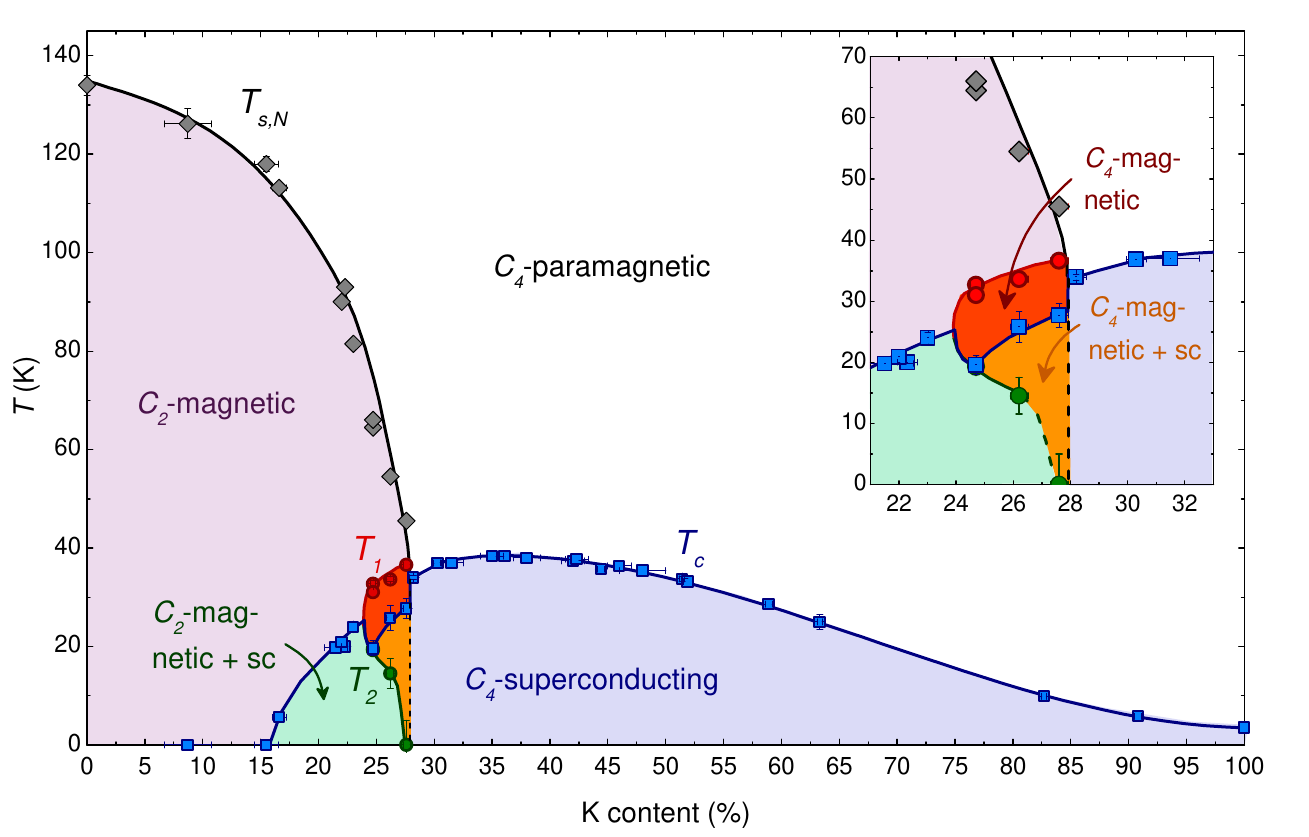}
\caption{\small {Detailed phase diagram of Ba$_{1-x}$K$_{x}$Fe$_2$As$_2$. (a) Phase diagram of Ba$_{1-x}$K$_{x}$Fe$_2$As$_2$ spanning the whole substition range from 0\%-100\% K content. Transition temperatures were obtained from thermodynamic measurements (thermal expansion of specific heat) while the K content was determined from refinement of 4-circle single-crystal x-ray diffraction patterns. Five distinct thermodynamically ordered phases are indicated by coloured areas. In particular, a narrow region of a $C_4$-symmetric (tetragonal) magnetic phase, occurring within the usual $C_2$-symmetric (orthorhombic) SDW phase, is observed. It coexists with superconductivity but suppresses $T_c$, while superconductivity-induced reentrance of the $C_2$ SDW phase occurs at $T_2$. The inset shows an enlarged view of the region containing the $C_4$ magnetic phase.}}
\label{fig:3-inset}
\end{figure*}

Increasing the K-content by just over 1\% (to 26.2\% K) drastically changes the behaviour. Superconductivity at $T_c=26$ K can now easily be identified by the clear second-order specific-heat anomaly and a small kink in $L_{ab}$ (Fig. 2 h,m). The specific-heat anomaly is again broadened and shifted to lower temperatures by a high magnetic field. Superconductivity now coexists with the intermediate $C_4$ phase and reentrance of the $C_2$ phase occurs at $T_2<T_c$, as evidenced by the increase of $\delta$ and $L_{ab}$ (Fig. 2 c,h). Surprisingly, there is only a very small anomaly in the heat capacity at $T_2$ (Fig. 2 m). At still a slightly higher K content (27.6\% K, Fig. 2 d,i,n,s), no more anomaly that could be associated with $T_2$ is observed in either the heat capacity or in $L_{ab}$. The weak re-emergence of the orthorhombic distortion at low temperature is likely induced by the stress applied for detwinning. For this sample, the reduction of $S_e/T$ is mainly due to superconductivity, and the magnetic and structural phase transitions at $T_{s,N}$ and $T_1$ play only a very minor role. Finally, Fig. 2 e,j,o,t show the results for a sample with 30\% K content, which undergoes only a superconducting transition. Note the larger specific-heat anomaly, which implies a considerably larger superconducting condensation energy than for the other samples.

Another particularly striking feature of our heat-capacity data is the marked low-temperature contribution to the electronic specific heat, or equivalently to the entropy, in the superconducting state for samples with 26.2\% and 27.6\% K content (see Fig. 2 m,n,r,s). This feature, which is very reminiscent of the very small superconducting gap found in KFe$_2$As$_2$\cite{Hardy2014}, is a sign of excited quasiparticles far below $T_c$ and seems to occur only when the structural-magnetic transitions are weak, i.e., induce only a small entropy change. From the position of the maximum in $C_e/T$, we estimate for the size of the smallest superconducting gap $\Delta_{sc}\sim 0.07 k_{B} T_c$ in the multigap system, which is even smaller than the ``lilliputian'' gaps in KFe$_2$As$_2$\cite{Hardy2014}. The occurrence of such an extremely small gap may be related to peculiar features of the Fermi surface resulting from a reconstruction at $T_{s,N}$ and $T_1$. When this reconstruction becomes weaker upon K doping, some parts of the reconstructed Fermi surface may move to within the superconducting gap $\Delta_{sc}$ of the Fermi level and can contribute to the superconducting condensate, as recently argued by Koshelev et al. \cite{Koshelev2014}. This would explain why this low-temperature feature suddenly disappears once $T_{s,N}$ and $T_1$ are suppressed by doping (see Fig. 2 o,t).  

The transition temperatures from Fig. 2 are summarized in the phase diagram of Fig. 3 together with additional thermodynamic data covering the whole phase diagram \cite{Boehmerthesis,Hardyunpublished}. We find five distinct thermodynamic ordered phases, which all compete for the electronic entropy provided by the high-temperature $C_4$-paramagnetic phase: $C_2$-SDW, $C_4$-magnetic, $C_2$ SDW coexisting with superconductivity , $C_4$-magnetic coexisting with superconductivity and $C_4$-superconducting. In the following, we discuss this competition in more detail. First, we note that there is a large sudden drop in the $T_c$ value of about 7 K as one goes from the $C_2$ to the $C_4$ magnetic phase. This points to a much stronger competition between superconductivity and the $C_4$-magnetic phase than between superconductivity and the $C_2$-magnetic phase, which may be due to the additional pronounced suppression of entropy at $T_1$ (see Fig. 2 q).
In particular, it would be clearly thermodynamically advantageous for superconductivity if the system would revert back to the $C_2$-magnetic phase with the higher electronic entropy available for pairing. The system apparently does just this via a peculiar first-order magnetic, structural and superconducting transition at $T_2$ for the crystal with 24.7\% K content. Upon further K doping, $T_1$ increases slightly, even though the entropy reduction at $T_1$ becomes significantly weaker. Apparently, this renders the coexistence of superconductivity and the magnetic $C_4$ phase possible, diminishing the driving force of the reentrant transition at $T_2$. Finally, $T_c$ suddenly increases by about 6 K at the boundary from the $C_4$-magnetic to the $C_4$-paramagnetic state. This can also be understood in terms of competing superconducting and magnetic order parameters, where the abrupt increase of $T_c$ is a consequence of the  first-order-like suppression of magnetism upon K substitution \cite{Kang2014}.
   
\begin{figure}
\includegraphics[width=\textwidth]{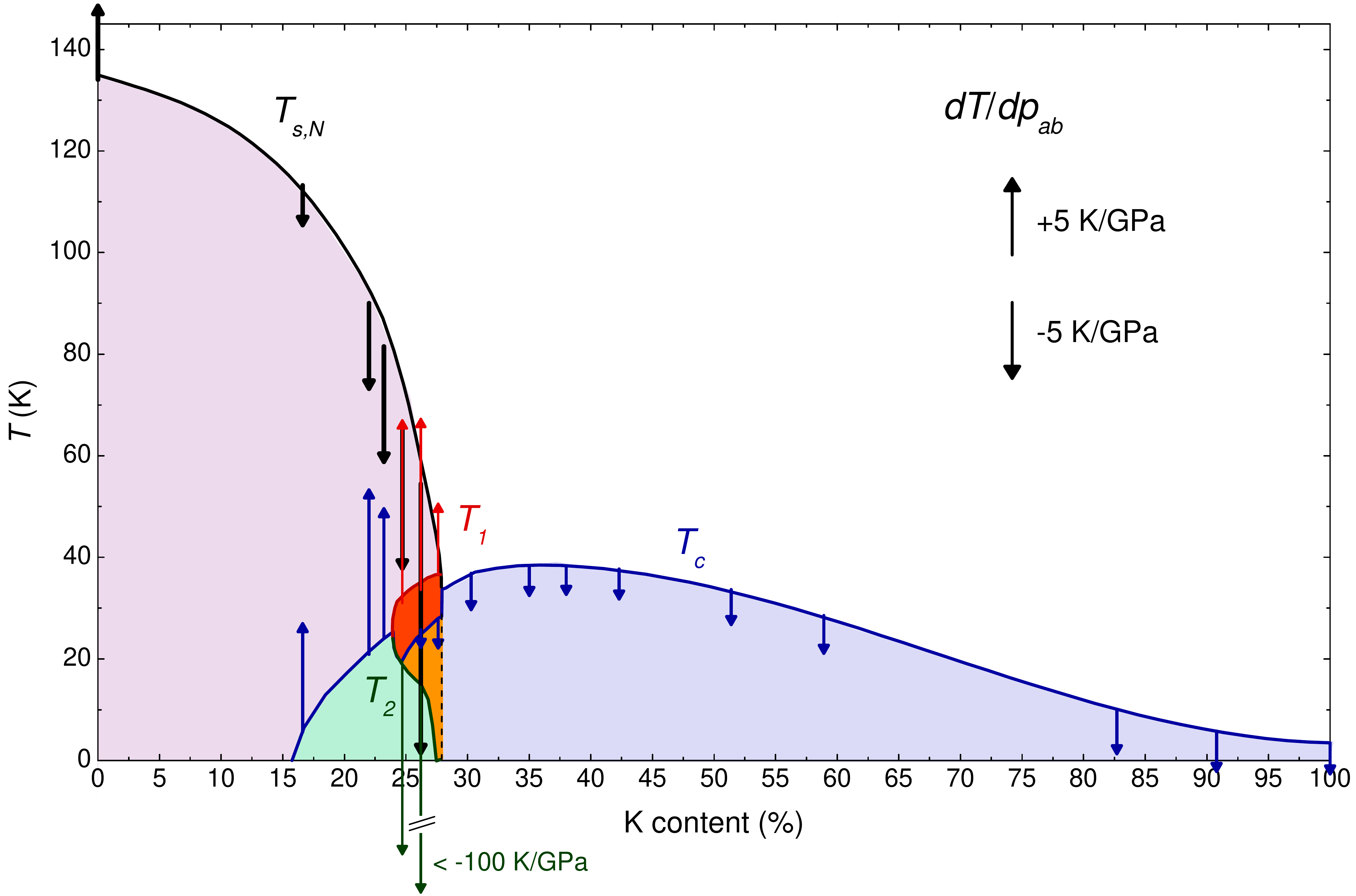}
\caption{\small {Effect of uniaxial in-plane pressure on the phase diagram of Ba$_{1-x}$K$_{x}$Fe$_2$As$_2$. The phase diagram of Fig. 3 is shown by continuous lines and vertical arrows represent the predicted effect of uniaxial in-plane pressure $p_{ab}$, determined from thermodynamic relations using our thermal-expansion and heat-capacity data. The direction of the arrows indicates whether the respective transition temperature increases or decreases and the length is proportional to the absolute value of the pressure derivatives (see scale).}}
\label{fig:4}
\end{figure}

The $C_4$ state which we observe in K-doped BaFe$_2$As$_2$ is very reminiscent of the magnetic $C_4$ phase observed in Na-doped BaFe$_2$As$_2$ \cite{Avci2014}, suggesting similar magnetic phases. We note that our macroscopic data in Fig. 2 (b) imply that 100\% of our Ba$_{1-x}$K$_{x}$Fe$_2$As$_2$ sample enters the $C_4$ magnetic state, while only roughly one half of the sample was found to transform to the new state in the polycrystalline samples of Avci et al \cite{Avci2014}. This difference is probably related to the much higher sample quality of the present single crystals. The most striking difference to (Ba,Na)Fe$_2$As$_2$ is, however, the sudden reemergence of $\delta$, i.e., re-entrance of the $C_2$ SDW phase, below  $T_2 \leq T_c$ in Ba$_{1-x}$K$_{x}$Fe$_2$As$_2$. One explanation for this difference may be a higher stability of the $C_4$ phase, indicated by considerably higher values of $T_1$ and a larger region of the $C_4$ phase in the Na-doped system. As pointed out by several authors \cite{Avci2014,Khalyavin2014,Wang2014,Wang2014II,Kang2014}, the detailed nature of the $C_4$-magnetic phase may hold important clues regarding the spin-orbital debate. Magnetic neutron scattering results on (Ba,Na)Fe$_2$As$_2$ single crystals point to a re-orientation of the spins from in-plane to $c$-axis orientated in the $C_4$ phase\cite{Wasser2014}, indicating that spin-orbit coupling cannot be neglected. There is also very likely a close link between our results and an additional phase observed in resistivity measurements under pressure in Ba$_{1-x}$K$_{x}$Fe$_2$As$_2$ (16\%-21\% K content) \cite{Hassinger2012} (see our resistivity measurement in the supplementary material), which were interpreted in terms of a second SDW transition. Even though it is impossible from our macroscopic measurements to determine the microscopic details of the magnetic ordering, we can nevertheless address the nature of these phase transitions from a thermodynamic point of view. In particular, the large entropy jump at the $T_1$ transition is not expected for a simple spin reorientation. Rather, our data imply that a significant change of the band structure occurs, which may be more in line with the proposed double-$Q$ magnetic structure in the itinerant spin-nematic viewpoint \cite{Avci2014,Kang2014}. 

Finally, we discuss the hydrostatic and uniaxial pressure dependences of the transition temperatures in our phase diagram (Fig. 4), which we determine from our thermal-expansion and specific-heat data using thermodynamic relations (see Methods) and which illustrate the phase competition from another point of view. The predicted effect of uniaxial (in-plane average) pressure $p_{ab}$ on the phase diagram of Ba$_{1-x}$K$_{x}$Fe$_2$As$_2$ is presented in Fig. 4. From 15\% to 24\% K content, superconductivity coexists with the $C_2$ SDW phase, and the pressure dependences $dT_{s,N}/dp_{ab} < 0$ and $dT_c/dp_{ab} > 0$ are of opposite sign and rather large due to the well-known competition between these two phases. Interestingly, a similar competition between the two magnetic phases is implied by the derivatives $dT_{s,N} /dp_{ab} < 0$ and $dT_1 /dp_{ab} > 0$. In particular, in-plane pressure is expected to strongly increase the extent of the C$_4$-magnetic phase due to the particularly large values of $dT_1/dp_{ab}$ and $dT_2/dp_{ab}$. This is similar to the observation that the $C_4$ phase is stabilized under hydrostatic pressure \cite{Hassinger2012}. Another notable feature is that, when $T_c$ lies within the $C_4$ magnetic phase, its pressure derivative, $dT_c/dp_{ab}\approx -1$ K/GPa, is much smaller than in the other phases, even though there apparently is a strong interaction between superconductivity and the magnetism in this phase (see Fig. 2). Note that the strong competition between superconductivity and the $C_2$ SDW phase manifests itself by a strong coupling of both order parameters to the orthorhombic distortion, and clearly such a coupling is not possible in a tetragonal state, which may explain the small values of $dT_c/dp_{ab}$. Finally, it is  of interest to note that $dT_c/dp_{ab}\approx -(2-3)$ K/GPa changes only weakly with doping within the whole range from 28\% to 100\% K content, advocating for no drastic change of the superconducting state over the whole phase diagram.

In summary, a new and very detailed phase diagram of Ba$_{1-x}$K$_{x}$Fe$_2$As$_2$, revealing intriguing new interactions between superconductivity and magnetism, has been derived from thermodynamic measurements. In particular, we find evidence for a narrow region of a $C_4$-symmetric magnetic phase which competes more strongly with superconductivity than the $C_2$-symmetric SDW phase. This competition between the two magnetic phases and superconductivity for the electronic entropy of the system results in a novel re-entrance of the $C_2$-symmetric phase either below $T_c$ or as a peculiar first-order concomitant superconducting, structural and magnetic transition. The strong reduction of entropy associated with it suggests this phase to be a $C_4$-type SDW in an itinerant scenario, rather than a simple spin re-orientation transition. However, there are still many unresolved issues, including the exact magnetic structure and the role of orbitals in this C$_4$ SDW phase. Its discovery at ambient pressure in high-quality single crystals opens up the possibility for studies using a large variety of more microscopic probes, such as magnetic neutron scattering, nuclear magnetic resonance, and angle-resolved photoemission studies, which will hopefully unravel some of these mysteries.  Finally, we note that our preliminary thermodynamic studies on the (Ba,Na)Fe$_2$As$_2$ system show that also this system is more complicated than initially thought, and a detailed comparison between the K- and Na-doped systems will be very useful \cite{Wangunpublished}.
 
\subsection{Methods}

High-quality samples were grown by a self-flux technique in alumina crucibles sealed in iron cylinders using very slow cooling rates of 0.2-0.4 $^\circ$C/hour. They were \textit{in situ} annealed by further slow cooling to room temperature. Single crystals with a mass of 1-3 mg were chosen for measurement and their composition was determined by refinement of 4-circle single-crystal x-ray diffraction patterns of a small piece of each crystal. Good homogeneity is attested by the sharp thermodynamic transitions.

Uniaxial thermal expansion was measured in a home-made capacitance dilatometer \cite{Meingast1990} and specific heat in a Physical Property Measurement System from Quantum Design. The electronic specific heat was obtained by subtracting \cite{Hardyunpublished}, from the raw data, a concentration-weighted sum of the lattice heat capacity of KFe$_2$As$_2$ and Ba(Fe$_{0.85}$Co$_{0.15}$)$_2$As$_2$ derived from Refs \cite{Hardy2014} and \cite{Hardy2010}. 

Uniaxial-pressure derivatives were obtained from the Clausius-Clapeyron relation $dT/dp_{ab}=V_m\left(\Delta L_{ab}/L_{ab}\right)/\Delta S$ for first-order phase transitions ($T_{s,N}$, $T_1$, $T_2$) and from the Ehrenfest relation, $dT/dp_{ab}=V_m\Delta\alpha_{ab}/\Delta (C/T)$ for second-order phase transitions ($T_c$). Here, $p_{ab}$ is average in-plane pressure, $\alpha_{ab}=d(\Delta L_{ab}/L_{ab})/dT$ the thermal expansion coefficient and $\Delta L_{ab}$, $\Delta S$ and $\Delta\alpha_{ab}$ are the discontinuities at the phase transition and $V_m=61.4$ cm$^3$/mol is the molar volume, which hardly changes upon K subsitution. 

\subsection{Acknowledgements}
We thank C. Bernhard, A. Chubukov, R. Fernandes, E. Hassinger, A. Kreyssig, J. Schmalian, Y. Su and M. Tanatar for fruitful and stimulating discussions.


\end{document}


\preprint{APS/123-QED}

\date{\today}

\title{Superconductivity-induced reentrance of orthorhombic distortion in Ba$_{1-x}$K$_x$Fe$_2$As$_2$}

\author{A. E. Böhmer}
\email{present address: Ames Laboratory/ Iowa State University, boehmer@ameslab.gov}
\affiliation{Institut für Festkörperphysik, Karlsruhe Institute of Technology, 76021 Karlsruhe, Germany}

\author{F. Hardy} 
\affiliation{Institut für Festkörperphysik, Karlsruhe Institute of Technology, 76021 Karlsruhe, Germany}

\author{L. Wang} 
\affiliation{Institut für Festkörperphysik, Karlsruhe Institute of Technology, 76021 Karlsruhe, Germany}

\author{T. Wolf}
\affiliation{Institut für Festkörperphysik, Karlsruhe Institute of Technology, 76021 Karlsruhe, Germany}

\author{P. Schweiss}
\affiliation{Institut für Festkörperphysik, Karlsruhe Institute of Technology, 76021 Karlsruhe, Germany}

\author{C. Meingast}
\email{christoph.meingast@kit.edu}
\affiliation{Institut für Festkörperphysik, Karlsruhe Institute of Technology, 76021 Karlsruhe, Germany}
\maketitle
\section{Supplemental material}

\subsection{Measurement of orthorhombic distortion}

At the characteristic tetragonal-to-orthorhombic phase transition of the iron-based materials, the lattice expands along one diagonal of the tetragonal high-temperature unit cell (which then becomes the orthorhombic $a$ axis) and shrinks along the other diagonal (subsequently the orthorhombic $b$ axis). $\mu$m-sized structural domains, or 'twins', in which orthorhombic $a$ and $b$ axes are interchanged with respect to each other, are distributed approximately evenly within the sample \cite{Tanatar2009}, unless a symmetry-breaking uniaxial pressure (stress) is applied. Uniaxial pressure along the tetragonal [110] direction  'detwins' the sample such that, ideally, only one type of domains with the (shorter) orthorhombic $b$ axis aligned along the direction of the applied pressure remains \cite{Fisher2011}.

\begin{figure*}
\includegraphics[width=17.2cm]{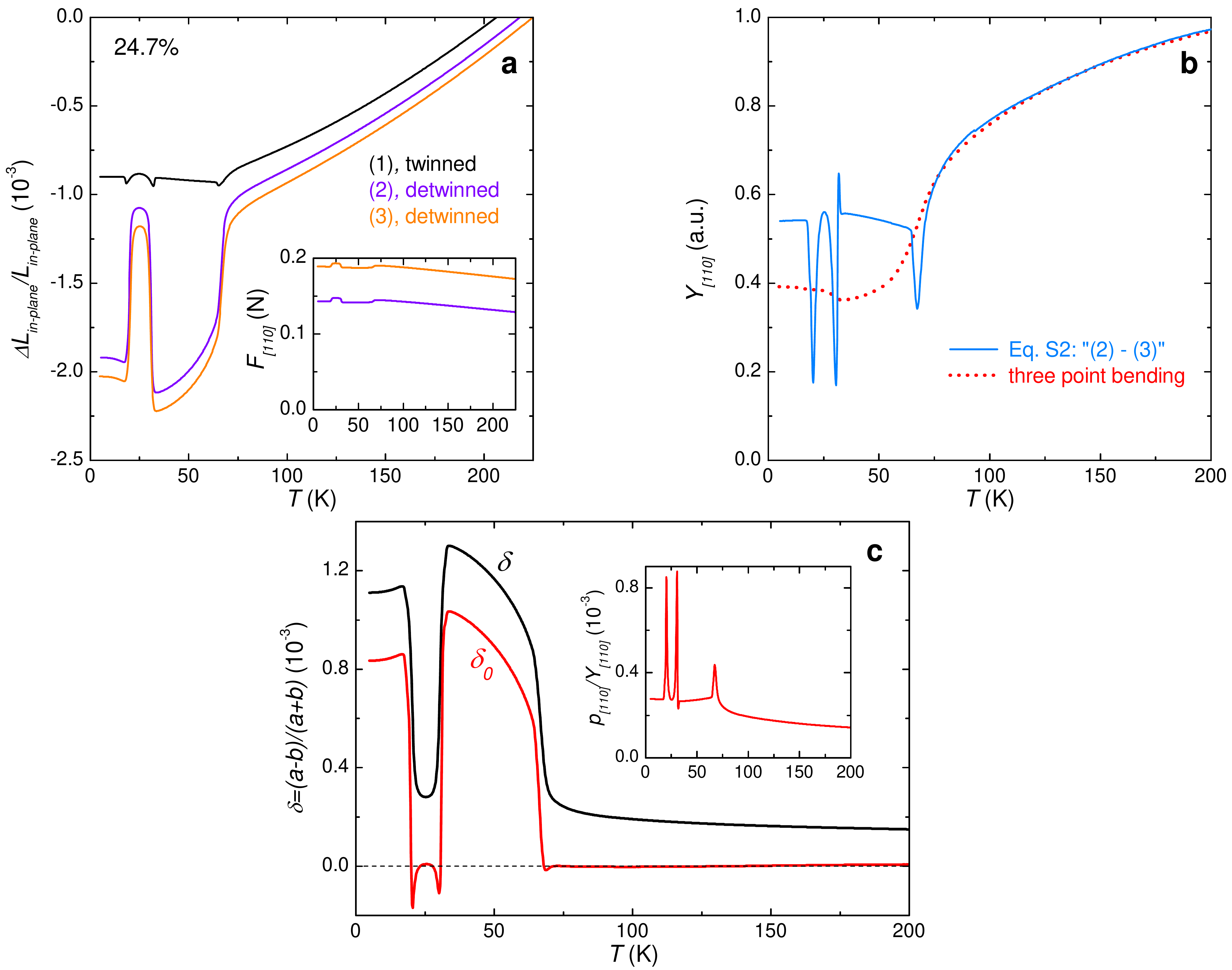}
\caption{Determination of the orthorhombic distortion using thermal expansion measurements under varying load in a capacitance dilatometer. (a) In-plane length change of the sample with 24.7\% K content under varying force $F_{[110]}$ applied by the spring-loaded dilatometer. In the 'twinned' measurement (1), (black curve) the force was applied along [100], hence $F_{[110]}=0$. The values of $F_{[110]}$ applied during measurements (2) and (3) are given in the inset. (b) Young's modulus $Y_{[110]}$ obtained from the data in (a) using equation \ref{eq:parallelplate}. The dotted line shows the result of a three-point bending experiment in a capacitance dilatometer \cite{Boehmer2014}. Curves in (a) were shifted for the $Y_{[110]}$ to match at high temperatures (see main text). (c) Orthorhombic distortion $\delta$ obtained by taking the difference between 'twinned' dataset (1) and 'detwinned' dataset (3) from panel (a) (blue curve). Data corrected for the elastic effect of the applied force, $\delta_0$ (red curve), by subtracting the induced elastic length change $p_{[110}]/Y_{[110]}$ shown in the inset.}
\label{fig:S1}
\end{figure*}

Our measurements of the orthorhombic distortion $\delta=(a-b)/(a+b)$ in the capacitance dilatometer \cite{Meingast1990} rely on the fact that the spring-loaded dilatometer cell intrinsically applies a small force $F$ on the sample along direction of the measurement. This force is high enough to detwin the crystal when a sample is inserted such that it is directed along the tetragonal [110] direction and thus the thermal expansion of the orthorhombic $b$ axis is obtained. In contrast, when the sample is inserted along the tetragonal [100] direction, the sample remains twinned and an average of orthorhombic $a$ and $b$ axis is measured. The orthorhombic distortion can then be computed by taking the difference of the 'twinned' and the 'detwinned' data. Note that, if a sample is not completely detwinned or, conversely, partly detwinned in the nominally 'twinned' measurement, $\delta$ will be underestimated but the temperature dependence will hardly be affected.

It is important to realize that the applied force at all temperatures also induces an elastic change of the sample length, simply according to Hooke's law. This 'additional' length change is, hence, given by 
\begin{equation}
\Delta L_{[110]}/L_{[110]}=S_{[110]}p_{[110]},
\end{equation}
 where $\Delta L_{[110]}/L_{[110]}$ is the relative length change of the sample along [110], $p_{[110]}$ is the uniaxial pressure on the sample generated by the dilatometer force $F_{[110]}$ and $S_{[110]}=Y^{-1}_{[110]}$ is the sample's elastic compliance (inverse Young's modulus) along the tetragonal [110] direction. If $Y_{[110]}$ were temperature independent, the elastic effect would result in a constant contribution to $\delta$. However, $Y_{[110]}$ of underdoped (Ba,K)Fe$_2$As$_2$ is strongly temperature dependent \cite{Boehmer2014}. To quantify the effect of the applied force, a bar-shaped sample (24.7\% K content) of dimensions  $(3.066\times1.08\times0.060)$ mm$^3$ with the longest dimension along [110] and the shortest dimension along the $c$ axis  has been measured under varying $F_{[110]}$ (measurements (2) and (3) in Fig. S1 a). The applied force, which is varied by changing the initial gap of the capacitor, was determined accurately and is slightly temperature dependent (inset in Fig. 1 a). An additional 'twinned' measurement (1) was conducted on another piece of the same larger crystal oriented along [100] (black curve in Fig. S1 a). We assume that the two types of domains are evenly distributed in the 'twinned' measurement (1) and that the sample is completely detwinned in the measurements (2) and (3). 

In principle, the Young's modulus can then be obtained by taking the difference of two measurements as 
\begin{equation}
Y_{[110]}=\frac{p_{[110]}^{(n)}-p_{[110]}^{(m)}}{L_{[110]}^{(n)}-L_{[110]}^{(m)}},
\label{eq:parallelplate}
\end{equation}
where $p_{[110]}$ is the applied uniaxial pressure and the superscripts $(n)$ and $(m)$ stand for measurements with different values of the applied force. A complication arises because $L_{[110]}^{(n)}-L_{[110]}^{(m)}$ is determined only up to a constant in the dilatometer. This constant can, however, be determined by shifting the curves in Fig. S1 a vertically with respect to each other so that the Young's modulus obtained via eq. \ref{eq:parallelplate} matches the Young's modulus measured in a three-point bending experiment \cite{Boehmer2014}. The advantage of using eq. \ref{eq:parallelplate} over three-point bending is that $Y_{[110]}$ can be obtained over the whole temperature range. In contrast, the bending is strongly affected by the presence of structural twins \cite{Schranz2012}, so that 'intrinsic' monodomain behavior cannot be obtained in the orthorhombic state. Note that, in order to get agreement between the Young's moduli from the two measurement techniques, the curves (2) and (3) are shifted by $0.065\times10^{-3}$ with respect to each other at $T=150$ K, which corresponds to a reasonable value of $Y_{[110]}(150\textnormal{ K})\approx70$ GPa, when assuming that roughly $1/5$ of the sample cross-section is in direct contact with the dilatometer cell. The so obtained $Y_{[110]}$ (Fig. 1 b) exhibits the expected softening at the three structural phase transitions at $T_{s,N}$ $T_1$ and $T_2$ and is harder between these temperatures. 

\begin{figure*}
\includegraphics[width=17.2cm]{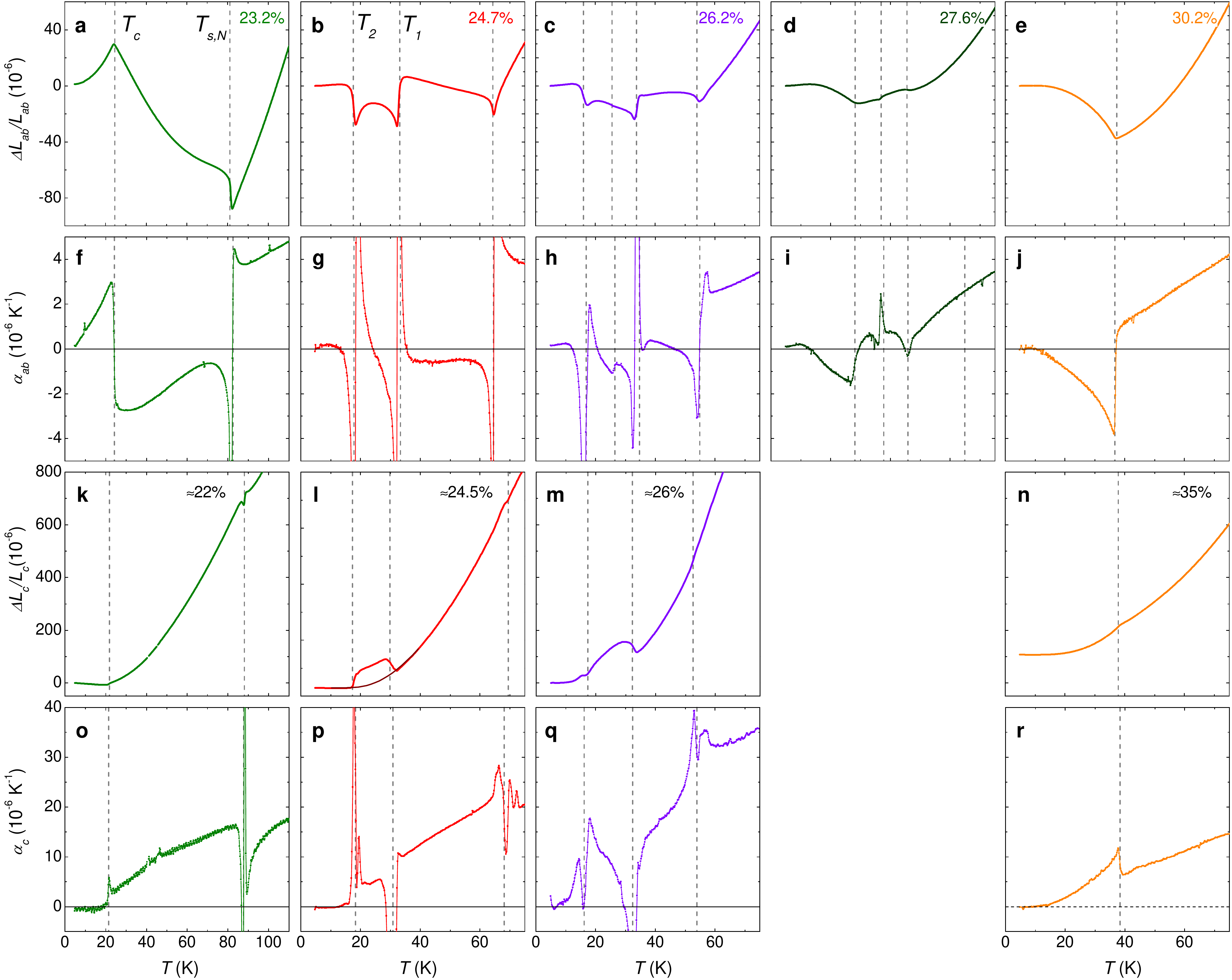}
\caption{Uniaxial thermal expansion of (Ba,K)Fe$_2$As$_2$ for $\sim23\%-30\%$ K content. (a-e) In-plane length change, as in Fig. 2 a-e of the main article. (f-j) Its temperature derivative, the uniaxial thermal-expansion coefficient $\alpha_{ab}$, highlighting smaller anomalies at $T_c$. (k-n) $c$-axis length change of samples with similar composition. The thin line in (l) supports reentrance of the orthorhombic spin-density-wave phase at $T_2$. (o-r) The $c$-axis thermal-expansion coefficient $\alpha_c$.}
\label{fig:S3}
\end{figure*} 

Using these results, we can, finally, quantify how the measured orthorhombic distortion is affected by the stress applied for detwinning. For example, in the above measurement, $F_{[110]}=0.18$ N induces a distortion of $\delta=0.16\times 10^{-3}$ already at 150 K, i.e., far above the structural transition. The 'intrinsic' distortion corrected for the effect of applied stress, $\delta_0$, is obtained by subtracting the elastically induced contribution $p_{[110]}/Y_{[110]}$ (inset in Fig. S1 c). Indeed, we find that $\delta_0=0$ to within $\approx 0.01\times10^{-3}$ in both the tetragonal phases (Fig. S1 c and Fig. 1 f), which shows that the procedure works well. Only in the immediate vicinity of the phase transitions, unphysical negative values of $\delta_0$ are obtained, presumably because the linear stress-strain relationship of eq. \ref{eq:parallelplate} overestimates the effect of the applied stress in these regions. 

\section{Uniaxial thermal expansion}

Fig. S2 shows additional thermal-expansion data for the samples in Fig. 2 of the main article and similar samples. The uniaxial in-plane thermal-expansion coefficient $\alpha_{ab}=d(\Delta L_{ab}/L_{ab})/dT$ is helpful to locate small anomalies. In particular, the small kink in $L_{ab}$ at $T_c$ of the 26.2\% sample is clearly seen as a discontinuity in $\alpha_{ab}$ (see Fig. S2 h). The size of the jump at $\alpha_{ab}$ at $T_c$ is used to compute the pressure derivative of $T_c$ shown in Fig. 4 of the main article. Fig. S3 k,l,m,n show the $c$-axis length changes $\Delta L_c/L_c$ for samples of similar composition. Note that, in order to get reliable data, ``thicker'' samples, i.e., samples which are longer along the $c$ axis, had to be chosen and these often have a slightly more inhomogeneous K content, resulting in broadened anomalies. Nevertheless, $T_{s,N}$, $T_1$ and $T_2$ can clearly be defined also in the $c$-axis data for samples with $\sim24.5\%$ and $\sim26\%$ K content. In particular, an extrapolation of the $\Delta L_c/L_c$-data supports that re-entrance of the original $C_2$ SDW state occurs below $T_2$ (see dashed line in Fig. S3 l). In general, in the anomalies at the phase transitions in the $c$-axis data are smaller with respect to the ``background'' expansion , while they have opposite sign and similar shape with respect to the in-plane data. This seems to be a quite general feature of the iron-based superconductors \cite{Hardy2009,Meingast2012}.

\section{Resistivity measurements}

Fig. S4 shows the in-plane normalized electrical resistivity of a crystal with $\approx 25\%$ K content. The transitions at $T_{s,N}$ and $T_1$ are sharp and well-defined  and closely match the thermodynamic data for samples of similar composition. Importantly, the transition at $T_1$ is very reminiscent of the unidentified transition at ``$T_0$'' observed in underdoped (Ba,K)Fe$_2$As$_2$ under pressure in Ref. \onlinecite{Hassinger2012}, suggesting that the same phase transition is observed also under pressure. The re-entrance of the $C_2$ SDW phase at $T_2$ within the superconducting state is, of course, impossible to observe using resistivity data. We note that the drop in resistivity at $T_c$ is quite broad and is significantly higher than the thermodynamically determined $T_c$ value, indicating that that  resistivity is not a good measure of the bulk superconducting transition.

\begin{figure}
\includegraphics[width=8.6cm]{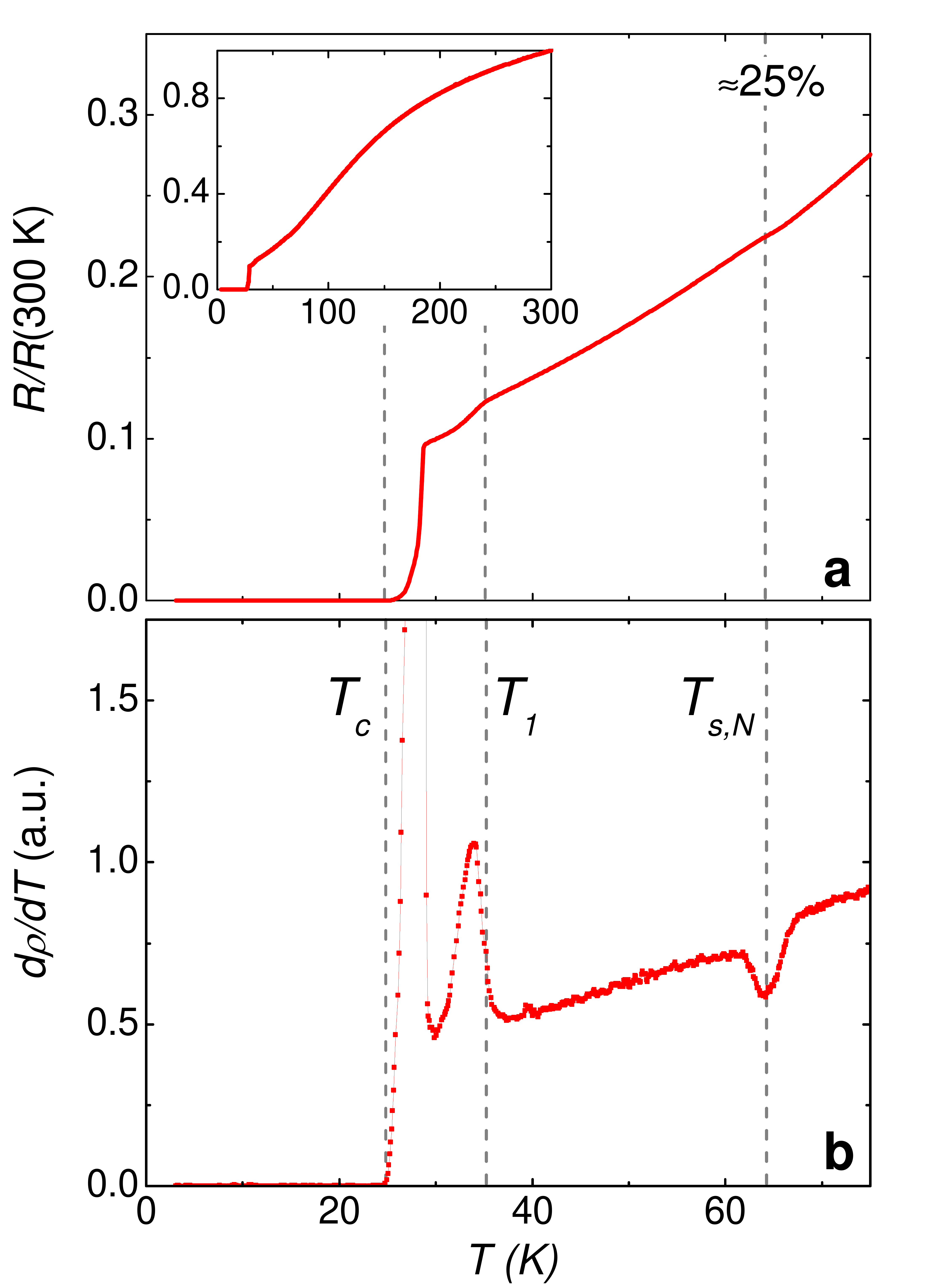}
\caption{Electrical resistance of a sample with $\approx25\%$ K content. (a) Electrical resistance normalized at room temperature. The inset shows the data over the whole temperature range. (b) Temperature derivative $d\rho/dT$ clearly showing sharp transitions at $T_{s,N}$ and $T_1$.}
\label{fig:S4}
\end{figure}
